\documentclass[amsmath,amssymb]{epl}
  
\begin{document}

\title{Scale-free Networks from Optimal Design}
 
\author{S. Valverde\inst{1}, R. Ferrer Cancho\inst{1} \and R. V. Sol\'{e}\inst{1,2}}
\institute{
   \inst{1} ICREA-Complex Systems Lab, IMIM-UPF, Dr Aiguader 80, Barcelona 08003, SPAIN \\
   \inst{2} Santa Fe Institute, 1399 Hyde Park Road, New Mexico 87501, USA.
}



\maketitle

\begin{abstract}
A large number of complex networks, both natural and artificial, share
the presence of highly heterogeneous, scale-free degree
distributions. A few mechanisms for the emergence of such 
patterns have been suggested, optimization not being one of
them. In this letter we present the first evidence for the emergence of scaling
(and the presence of small world behavior) 
in software architecture graphs from a well-defined local optimization
process. Although the rules that define the strategies involved in
software engineering should lead to a tree-like structure, the final
net is scale-free, perhaps reflecting the presence of conflicting
constraints unavoidable in a multidimensional optimization
process. The consequences for other complex networks are outlined.
\end{abstract}

Two basic features common to many complex networks, from the Internet
to metabolic nets, are their scale-free (SF) topology \cite{albertbarabasi} 
and a small-world (SW) structure \cite{wattsstrogatz,newmanSW}. The first states 
that the proportion of nodes $P(k)$ having $k$ links decays as a
power law $P(k) \sim k^{-\gamma}\phi(k/\xi)$ (with $\gamma \approx 2-3$)
\cite{albertbarabasi,barabasialbert,amaral} (here $\phi(k/\xi)$ introduces a cut-off 
at some characteristic scale $\xi$). Examples of SF nets include
Internet topology \cite{barabasialbert,caldarelli}, 
cellular networks \cite{jeongprot,jeongmet}, scientific
collaborations \cite{newmanpnas} and \cite{ferrersoleLANG}
lexical networks. The second refers to a web 
exhibiting very small average path lengths between nodes along with a 
large clustering\cite{wattsstrogatz,newmanSW}.

Although it has been suggested that these nets originate from preferential
attachment \cite{barabasialbert}, the success of theoretical 
approximations to branching nets from optimization
theory\cite{westbrown,rinaldo} would support optimality as an alternative
scenario. In this context, it has been shown that minimization of both vertex-vertex distance and
link length ({\em i.e.} Euclidean distance between 
vertices) \cite{mathias} can lead to the SW phenomenon. In a
similar context, SF networks have been shown to originate from a 
simultaneous minimization of link density and 
path distance\cite{ferrersoleoptim}. Optimal wiring has also been proposed within the
context of neural maps \cite{cherniak}: 'save wiring' is an organizing
principle of brain structure. However, although the analysis of
functional connectivity in the cerebral cortex has shown evidence for
SW \cite{stephan}, the degree distribution is clearly
non-skewed but single-scaled (i. e. $\xi$ is very small). 

The origin of highly heterogeneous nets is particularly important since it has been
shown that these networks are extremely resilient under random
failure: removal of randomly chosen nodes (tipically displaying low
degree) seldom alters the fitness of the net \cite{albertachiles}. 
However, when nodes are removed by sequentially eliminating 
those with higher degree, the system rapidly experiences network 
fragmentation\cite{albertachiles,solemonty}.

\begin{figure}
\begin{center}
\includegraphics[scale=0.6]{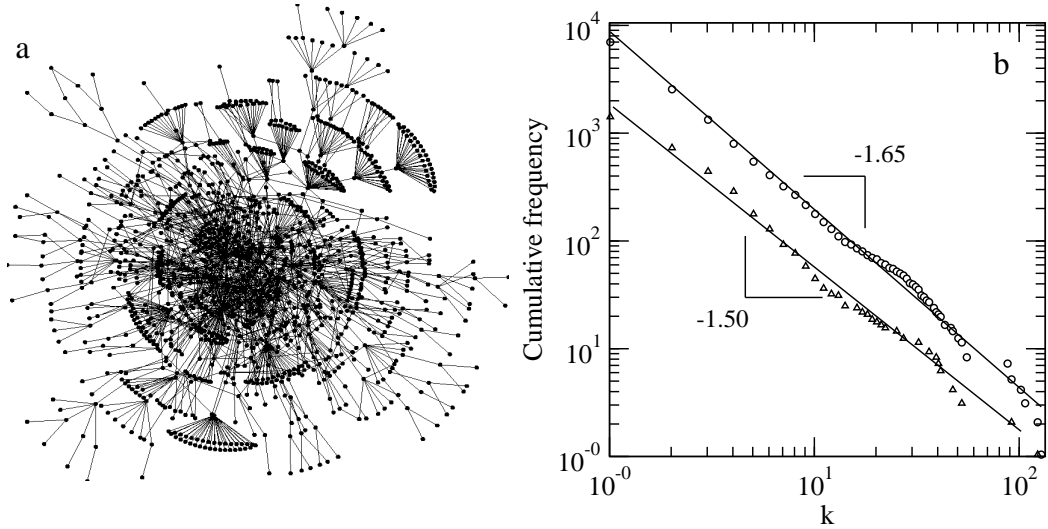}
\end{center}
\caption{\label{fig:EPLgrafs} (a) One of the largest components of the java
net ($\Omega_2$, displays scale-free and small world behavior (see
text). In (b) the cumulative frequencies $P_>(k)$ are shown for the two
largest components. We have $P_>(k)
\sim k^{-\gamma+1}$, with $\gamma_1=1.5 \pm 0.05$ and  $\gamma_2=1.65
\pm 0.08$.}
\end{figure}

Artificial networks offer an invaluable reference when dealing with
the rules that underlie their building process \cite{circuits}.
Here we show that a very important class of networks derived from 
software architecture maps, displays the previous patterns as a 
result of a design optimization process.  

The importance of software and understanding how to build efficiently 
software systems is one of our major concerns. Software is present in 
the core of scientific research, economic markets, military equipments
and health care systems, to name a few. Expensive costs (thousands of 
millions of dollars) are associated with the software development process.
In the past 30 years we have assisted to the birth and technological evolution of software 
engineering, whose objective is to provide methodologies and tools to
control and build software efficiently. Software engineers 
conceive programs with graphs as architects use 
plans for buildings. The software architecture is the structure of the program.
The building blocks are software components and links are relationships between 
software components. The interactions between all the components yields the program
functionality. {\em Class diagrams} constitute a well-known example 
of such graphs\cite{designpatterns}.
In this case, software components are also known by the technical term {\em class}. 
We have analysed the class diagram of the public Java Development
Framework 1.2 (JDK1.2) \cite{jdk12}, which is a large set of software components widely used by 
Java applications, as well as the architecture of a large computer game\cite{prorally}. 

These are examples of highly optimized structures, where design 
principles call for diagram comprehensibility, grouping components into modules , 
flexibility and reusability (i.e. avoiding the same task to be performed
 by different components) \cite{Pressman}.  Although the entire plan is controlled by 
engineers, no design principle explicitely introduces preferential 
attachment nor scaling and small-worldness. The resulting graphs,
however, turn out to be SW and SF nets.
 
The software graph is defined by a pair $\Omega_s=(W_s, E_s)$, where $W_s=\{ s_i
\}, (i=1, ..., N)$ is the set of $N=\vert \Omega \vert$ classes and $E_s=\{ \{s_i, s_j\} \}$
is the set of edges/connections between classes.  The {\em adjacency
  matrix} $\xi_{ij}$ indicates that an interaction exists between
classes $s_i, s_j \in \Omega_s$ ($\xi_{ij} = 1$) or that the interaction is
absent ($\xi_{ij} = 0$). The average path lenght $l$ is given by the average $l = \left <
l_{min}(i,j) \right >$ over all pairs $s_i, s_j \in \Omega_s$, where
$l_{min}(i,j)$ indicates the length of the shortest path between two nodes. 
The clustering coefficient is defined as the probability that two
classes that are neighbors of a given class are neighbors of each
other. Poissonian graphs with an average degree $\bar{k}$ are such
that $C \approx \bar{k}/N$ and the path length follows
\cite{newmanSW}: 
\begin{equation}
l \approx {\log N \over \log (\bar{k})}
\end{equation}
$C$ is easily defined from the adjacency matrix, and is given by:
\begin{equation}
C= \left < 
\frac{2}{k_i (k_i-1)} \sum_{j=1}^{N} \xi_{ij}  \left [ \sum_{k
      \in \Gamma_i} \xi_{jk} \right ]
\right >_{\Omega_s}
\end{equation}
It provides a measure of the average fraction of pairs of
neighbors of a node that are also neighbors of each other.
 
The building process of a software graph is done in parallel
(different parts are build and gradually get connected) and is assumed to follow some
standard rules of design \cite{Pressman,designpatterns}. None of these rules refer to the
overall organization of the final graph. Essentially, they deal with 
optimal communication among modules and low cost (in terms of wiring)
together with the rule of avoiding hubs (classes with large number 
of dependencies, that is, large degree). 
The set of bad design practices, such as making use of large hubs, is known as 
{\em antipatterns} in the software literature: see\cite{antipatterns}. The development 
time of the application should be as short as possible because the expensive costs 
involved. It is argued in literature 
\cite{Pressman} that there is an optimum number of components so that
cost of development is minimized, but it is not possible 
to make a reliable prediction about this number. 
Adding new software components involves more cost in terms of interconnections 
between them (links). Conversely, the cost per single software component decreases 
as the overall number of components (nodes) is increased because the 
functionality is spread over the entire system. Intuitively, a  trade-off between 
the number of nodes and the number of links must be chosen. 

However, we have found that this (local) optimization process
results in a net that exhibits both scaling and small-world
structure. First, we analyzed JDK1.2 network has $N=9257$ nodes 
and $N_c=3115$ connected components, so that the complete graph
$\Omega_s$ is actually given by $\Omega_s=\cup_{i} \Omega_i$, where
the set is ordered from larger to smaller components 
($\vert  \Omega_1 \vert > \vert  \Omega_2 \vert > ... > \vert
\Omega_{N_c} \vert$). The largest connected component, $\Omega_1$,  
has $N_1=1376$, with $<k>=3.16$ and $\gamma=2.5$, with 
clustering coefficient [4] is $C=0.06 \gg C^{rand}=0.002$ 
and the average distance $l=6.39 \approx l^{rand}=6.28$, i.e. it is a
small-world. 
The same basic results are obtained for $\Omega_2$ (shown in fig. 1a): here we have 
$N_2=1364$, $<k>=2.83$ and $\gamma=2.65$, $C=0.08 \gg C^{rand}=0.002$ 
and $l=6.91 \approx l^{rand}=6.82$. 

The degree distribution for the two largest components is shown in figure 1b, where we have
represented the cumulative distribution 
\begin{equation}
P_>(k; \Omega_i) = \sum_{k'\ge k}^{N(\Omega_i)} p(k', \Omega_i)
\end{equation}
for $i=1,2$. We can see that the largest components display scaling, with estimated
exponents $\gamma \approx 2.5-2.65$. 

Similar results have been
obtained from the analysis of a computer game graph \cite{prorally}. 
This is a single, complex piece of software which consists 
of $N=1989$ classes involving different aspects like: real-time computer 
graphics, rigid body simulation, sound and music playing, graphical user interface and
memory management. The software architecture graph for the game has a large connected
component that relates all subsystems. The cumulative degree 
frequency for the entire system is scale-free, with $\gamma =  2.85
\pm 0.11$. The network also displays SW behaviour: the clustering coefficient is 
$C=0.08 \gg C^{rand}=0.002$ and the average distance is $l=6.2$, close
to $l^{rand}=4.84$.

These results reveal a previously unreported global feature of
software architecture which can have important consequences in both
technology and biology. This is, as far as we know, the first 
example of a scale-free graph resulting from a local optimization
process instead of preferential attachment \cite{barabasialbert} or duplication-rewiring
\cite{solePROT,vazquez} rules. Since the failure of a
single module leads to system's breakdown, no global homeostasis
has been at work as an evolutionary principle, as it might have
occured in cellular nets. In spite of this, the
final structure is very similar to those reported from the analysis of 
cellular networks. Second, our results suggest that 
optimization processes might be also at work in the latest, as it has 
been shown to occur in transport nets \cite{westbrown}. 

\begin{figure}
\begin{center}
\includegraphics[scale=0.55]{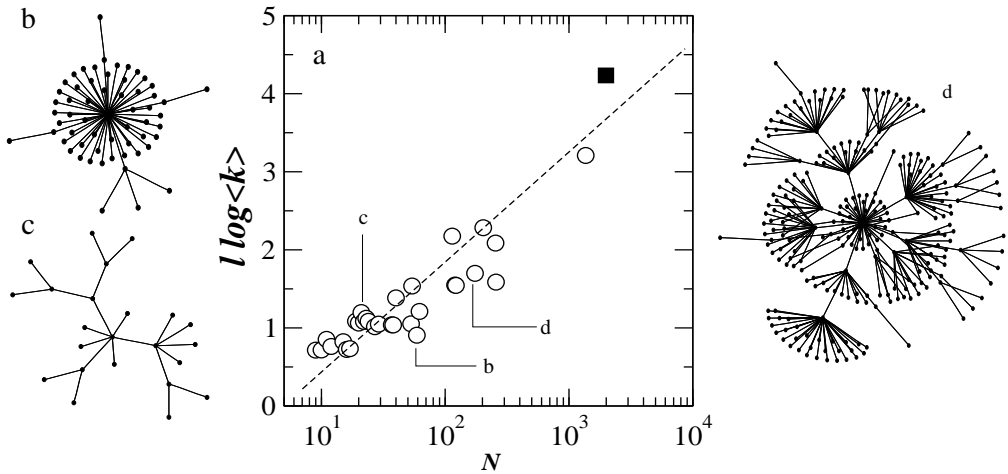}
\caption{\label{fig:properties_versus_lambda}(a) 
Using the $32$ connected components with more than 10 classes (nodes),
the $l \log(\bar{k})-N$ plots is shown. As predicted from a SW
structure, the components follow a straight line in this linear-log
diagram. Three subwebs are shown (c-d), displaying hubs but no
clustering (their location is indicated in (a)). The black square
corresponds to the computer game graph. }
\end{center}
\end{figure}

Complex biosystems are often assumed to result from selection
processes together with a large amount of tinkering \cite{jacob}. By contrast,
it is often assumed that engineered, artificial systems are highly
optimized entities, although selection would be also at work
\cite{monod}. Such differences should be observable when
comparing both types, but the analysis of both natural and artificial
nets indicates that they are often remarkably similar, 
perhaps suggesting general organization principles. Our results
support an alternative scenario to preferential attachment based on
cost minimization together with optimal communication among units
\cite{ferrersoleoptim} process. The fact that small-sized software graphs are
trees (as one would expect from optimization leading to hierarchical
structures, leading to stochastic Cayley trees \cite{caldarelli}) 
but that clustering emerges at larger sizes might be the
outcome of a combinatorial optimization process: As the number of modules
increases, the conflicting constraints that arise among different
parts of the system would prevent reaching an optimal structure
\cite{kauffman}. Concerning cellular networks, although preferential 
linking might have been at work \cite{wagnerfell}, optimization has probably
played a key role in shaping metabolic pathways
\cite{mittenthal,melendez1,melendez2}. We conjecture that the common origin of SF
nets in both cellular and artificial systems such as software might
stem from a process of optimization involving low cost (sparse graph)
and short paths. For cellular nets (but not in their artificial
counterparts) the resulting graph includes, for free, an enormous
homeostasis against random failure.

\begin{acknowledgments}
The authors thanks Javier Gamarra, Jose Montoya, William Parcher,
Charles Herman and Marcee Herman for useful comments. This work was 
supported by the Santa Fe Institute (RFC and RVS) and by grants of the Generalitat de 
Catalunya (FI/2000-00393, RFC) and the CICYT (PB97-0693, RVS). 
\end{acknowledgments}

\end{document}